\documentclass[11pt]{article}
\pdfoutput=1
\newcommand{\Comment}[1]{{}}
\usepackage{amsfonts,amsthm,amsmath,amssymb,slashed}
\usepackage[textwidth = 430 pt, textheight = 630 pt]{geometry}
\usepackage{color}

\Comment{\usepackage{color}
\definecolor{MyDarkBlue}{rgb}{0.15,0.15,0.45}
\usepackage[linktocpage=true]{hyperref}
\hypersetup{
colorlinks=true,
citecolor=MyDarkBlue,
linkcolor=MyDarkBlue,
urlcolor=MyDarkBlue,
pdfauthor={Horatiu Nastase and Marcelo R. Barbosa},
pdftitle={Penrose limit of MN solution and spin chains in three-dimensional field theories},
pdfsubject={hep-th}
}
\usepackage[utf8]{inputenc}
\usepackage[numbers,sort&compress]{natbib}
\usepackage{hypernat}}
\usepackage{graphicx}
\usepackage{cite}
\usepackage[colorlinks=true]{hyperref}
\hypersetup{
  allcolors = blue,
}

\newcommand\ignore[1]{}
\def\one{{\,\hbox{1\kern-.8mm l}}}

\def\Tr{{\rm Tr\, }}

\def\a{\alpha}\def\b{\beta}

\def\d{\partial}

\def\Tr{\mathop{\rm Tr}\nolimits}

\newcommand{\Cset}{{\,\,{{{^{_{\pmb{\mid}}}}\kern-.45em{\mathrm C}}}}}

\newcommand{\be}{\begin{equation}}
\newcommand{\bea}{\begin{eqnarray}}

\newcommand{\ee}{\end{equation}}
\newcommand{\eea}{\end{eqnarray}}

\begin{document}

\renewcommand{\thefootnote}{\fnsymbol{footnote}}

\makeatletter
\@addtoreset{equation}{section}
\makeatother
\renewcommand{\theequation}{\thesection.\arabic{equation}}

\rightline{}
\rightline{}




\begin{center}
{\LARGE \bf{\sc Penrose limit of MNa solution and spin chains in three-dimensional field theories}}
\end{center}
 \vspace{1truecm}
\thispagestyle{empty} \centerline{
{\large \bf {\sc Horatiu Nastase${}^{a}$}}\footnote{E-mail address: \Comment{\href{mailto:horatiu.nastase@unesp.br}}{\tt horatiu.nastase@unesp.br}}
{\bf{\sc and}}
{\large \bf {\sc Marcelo R. Barbosa${}^{b,}$}}\footnote{E-mail address: \Comment{\href{mailto:rezendemar@usp.br}}{\tt rezendemar@usp.br}}
                                                        }

\vspace{.5cm}

\centerline{{\it ${}^a$Instituto de F\'{i}sica Te\'{o}rica, UNESP-Universidade Estadual Paulista}}
\centerline{{\it R. Dr. Bento T. Ferraz 271, Bl. II, Sao Paulo 01140-070, SP, Brazil}}

\vspace{.3cm}

\centerline{{\it ${}^b$Instituto de F\'{i}sica, Universidade de S\~ao Paulo,}}
 \centerline{{\it Rua do Mat\~ao Travessa 1371, S\~ao Paulo 05508-090, SP, Brazil}}

\vspace{1truecm}

\thispagestyle{empty}

\centerline{\sc Abstract}

\vspace{.4truecm}

\begin{center}
\begin{minipage}[c]{380pt}
{\noindent 
In three dimensions, a theory with spontaneous breaking of supersymmetry was described holographically by the 
Maldacena-Nastase (MNa) solution. We revisit the issue of its Penrose limit, and compare the resulting pp wave and its string 
eigenstates to a sector of the theory on 5-branes reduced on $S^3$, and a spin chain-like system. We obtain many of the 
features of the pp wave analysis, but a complete matching still eludes us. We suggest possible explanations
for the resulting partial mismatch, and compare against other spin chains for 3-dimensional field theories with gravity duals.
}
\end{minipage}
\end{center}

\vspace{.5cm}

\setcounter{page}{0}
\setcounter{tocdepth}{2}

\newpage

\renewcommand{\thefootnote}{\arabic{footnote}}
\setcounter{footnote}{0}

\linespread{1.1}
\parskip 4pt


\tableofcontents

\section{Introduction}

The AdS/CFT correspondence \cite{Maldacena:1997re} and its gauge/gravity duality generalization
(see the books \cite{Nastase:2015wjb,Ammon:2015wua} for more details) 
gave a way to relate interesting, nonperturbative field theories to string theory in gravitational backgrounds, though string modes 
(as opposed to supergravity modes) were only obtained in the Penrose limit, leading to the pp wave correspondence 
\cite{Berenstein:2002jq}. The limit reduces the gravity dual to a very simple form, a pp wave, in which one can most often 
quantize the string and calculate its eigenmodes. On the field theory side, one focuses on a subset of operators of large dimension,
and a charge associated with a global symmetry, where again one can usually make some nontrivial calculation for the 
anomalous dimension. This makes it possible to understand better the holographic duality.

In four dimensions, the case one is most interested in because of phenomenology, one can consider the Penrose limit of the 
standard $AdS_5\times S^5$ vs. ${\cal N}=4$ SYM duality, as in the original paper \cite{Maldacena:2001pb}. 
A more interesting case would be of confining theories, like the Klebanov-Strassler \cite{Klebanov:2000hb} and 
Maldacena-N\'{u}\~nez \cite{Maldacena:2000yy} constructions. Their Penrose limits were first analyzed in \cite{Gimon:2002nr}, 
where the focus was on the IR limit of the solutions, and it was 
found that there the Klebanov-Strassler case gives a theory of hadrons dubbed "annulons" (for their ring-like structure), and the 
Maldacena-N\'{u}\~nez case was argued to be qualitatively similar, if harder to analyze.

In three dimensions, the standard conformal and ${\cal N}=6$ supersymmetric model is the ABJM model \cite{Aharony:2008ug}, 
and its Penrose limit was understood quite well in \cite{Nishioka:2008gz}. But three-dimensional theories turn out to be 
harder to understand. The Penrose limits of the GJV model \cite{Guarino:2015jca}, not yet a confining model, were analyzed in 
\cite{Araujo:2017hvi} and there were some puzzling mismatched quantities. 
The confining three-dimensional theory with dynamical supersymmetry breaking with a gravity dual, obtained from NS5-branes
wrapped on $S^3$ with a twist, was found by Maldacena and Nastase (MNa) \cite{Maldacena:2001pb}, and a start of an 
analysis for its Penrose limit was begun in \cite{Bertoldi:2004rn,Bigazzi:2004fc}, following the logic in 4 dimensions from 
\cite{Gimon:2002nr}, but again it was hard to describe in field theory the analog of the pp wave results. 

In this paper, we will revisit the Penrose limit of the MNa background and its field theory interpretation, and will try to do 
a more in depth analysis. We will find that most of the features of the pp wave analysis can be obtained, in particular 
the matching of the BPS states, but there are still mismatches in terms of the corrections to the spectrum and anomalous dimensions.
We suggest reasons for the mismatches, one of them 
being similar to the 4-dimensional case of Penrose limits of T-duals of $AdS_5\times S^5$, 
considered in \cite{Itsios:2017nou}. We will also compare these results to other results for gravity duals of three-dimensional 
systems, like the ABJM model, the GJV model, and the phenomenological holographic cosmology model of \cite{McFadden:2009fg}.

The paper is organized as follows. In section 2 we review the MNa solution and its field theory dual, in section 3
we find the pp wave in the Penrose limit for it, and quantize the string in this background. 
In section 4 we analyze the field theory modes and match them to the string modes on the pp wave, 
calculate corrections to the anomalous dimension and compare with the pp wave, then compare 
against other three dimensional cases previously analyzed. In section 5 we conclude and present some open questions.

\section{Review of MNa solution and dual field theory}

\subsection{Gravity dual}

The MNa solution is a solution of 10-dimensional type IIB string theory and 
was found by embedding a solution of 5-dimensional gauged supergravity by Chamseddine and Volkov
\cite{Chamseddine:2001hk} into 7 dimensions, and then into 10 dimensions. The solution for the metric, 
B field and dilaton is \cite{Maldacena:2001pb}
\bea
ds^2_{\rm 10, string}&=&d\vec{x}_{2,1}^2+\a' N\left[d\rho^2+R^2(\rho)d\Omega_3^2+\frac{1}{4}\left(\tilde w_L^a-A^a\right)^2\right]\cr
H=dB&=&N\left[-\frac{1}{4}\frac{1}{6}\epsilon_{abc}\left(\tilde w_L^a-A^a\right)\left(\tilde w_L^b-A^b\right)\left(\tilde w_L^c-A^c\right)
+\frac{1}{4}F^a\left(\tilde w_L^a-A^a\right)\right]+h\cr
h&=&N\left[w^3(\rho)-3w(\rho)+2\right]\frac{1}{16}\frac{1}{6}\epsilon_{abc}w_L^aw_L^bw_L^c\cr
A&=&\frac{w(\rho)+1}{2}w_L^a\cr
\Phi&=&\Phi(\rho)\;,\label{MNsol}
\eea
where $w^a_L$ and $w_R^a$ are the left- and right-invariant one-forms on $S^3$, respectively, 
where the $S^3$, with metric $d\Omega_3^2=w^a_Lw^a_L=w^a_Rw^a_R$, is the sphere on which the NS5-branes are wrapped, 
and $\tilde w^a_L$ and $\tilde w^a_R$ are the corresponding forms on the sphere at infinity (away from the NS5-branes), $S_\infty^3$. 
Specifically, we have (in a parametrization with angles $\psi, \theta,\phi$)
\bea
w_L^{1}  =\sin\psi d\theta-\sin\theta\cos\psi d\phi\;,&
w_L^{2}  =\cos\psi d\theta+\sin\theta\sin\psi d\phi\;,&
w_L^{3}  =d\psi+\cos\theta d\phi\cr
w_R^{1}  =-\sin\phi d\theta+\sin\theta\cos\phi d\psi\;,&
w_R^{2}  =\cos\phi d\theta+\sin\theta\sin\phi d\psi\;,&
w_R^{3}  =d\phi+\cos\theta d\psi\;,\cr
&&
\eea
and similar formulas for $\tilde w^a_L,\tilde w^a_R$ in terms of $\tilde\psi,\tilde \theta,\tilde \phi$, such that
\be
d\Omega_3^2=w_L^aw_L^a=d\theta^{2}+d\phi{}^{2}+d\psi^{2}+2cos\theta d\phi d\psi.
\ee

The functions $w(\rho),R^2(\rho)$ and $\phi(\rho)$ obey complicated coupled differential equations that can be solved perturbatively 
or numerically. The boundary condition (perturbative solution) in the UV is
\bea
R^2(\rho)&\simeq& \rho\;,\;\;\;
w(\rho)\simeq \frac{1}{4\rho}\;,\;\;\;
\Phi(\rho)\simeq -\rho+\frac{3}{8}\log \rho\;,\cr
\int_{S_\infty^3}h&=&k\;,
\eea
which corresponds to the gravity dual of little string theory. For the nonsingular solution with $k=N/2$, as above, 
supersymmetry is preserved. If one considers extra $\delta n =k-N/2$ NS5-branes wrapped on $S^3$, with $|k-N/2|\ll N$ so that 
there is no backreaction to gravity, $\delta n>0 $ (branes) keeps the solution supersymmetric, and $\delta n<0$ (antibranes) breaks 
supersymmetry, which is a dynamical breaking from the point of view of field theory.

The solution in the IR (near the origin $\rho=0$) is of more interest to us, and 
is (see also \cite{Chamseddine:2001hk}, \cite{Bertoldi:2004rn})
\bea
R(\rho)&\simeq& \rho-\frac{2+9b^2}{36}\rho^3+{\cal O}(\rho^5)\cr
w(\rho)&\simeq& 1-b\rho^2+{\cal O}(\rho^4)\cr
\Phi(\rho)&\simeq & \Phi_0-\frac{2+3b^2}{8}\rho^2+{\cal O}(\rho^4)\;,
\eea
where a parameter $b$ was introduced, which for the supersymmetric solution has the value $b=1/3$, and other values in $[0,1)$ 
correspond to soft breaking of supersymmetry by a gaugino mass term (see 
\cite{Bertoldi:2004rn,Bigazzi:2004fc,Bertoldi:2002ks,Gomis:2001xw} for these generalizations and the notation).

It turns out that for the description of the field theory, it is better to S-dualize the solution, to a solution for D5-branes wrapped on 
$S^3$ with a twist, with the string frame metric, H-field and dual dilaton in the IR 
\be
ds^2_{\rm 10, D, string}=e^{\Phi_D(\rho)}ds^2_{\rm 10, string}\;,\;\; H^{(D)}=e^{\Phi_D(\rho)}H\;,\;\;
\Phi_D(\rho)\simeq \Phi_{D,0}+\frac{2+3b^2}{8}\rho^2+{\cal O}(\rho^4).
\ee

\subsection{Field theory}

The field theory corresponds to the theory on NS5-branes on $S^3$, with a twist that preserves ${\cal N}=1$ supersymmetry. 
The spin connection on $S^3$ is in $SU(2)$, which is isomorphic to the coset $SO(4)/SO(3)=S^3$. The NS5-brane theory 
has an $SO(4)\simeq SU(2)_L\times SU(2)_R$ R-symmetry, corresponding in the gravity dual to the isometry of the 
sphere at infinity $S_\infty^3$, and the twisting amounts to embedding the spin connection into $SU(2)_L$, which was realized 
in the gravity dual by identifying the gauge field $A^a$, which takes value in the $SU(2)_L\subset SO(4)$ of the 
$S_\infty^3$ (coupled to R-symmetry), with the spin connection $w^a_L$ on the $S^3$, in the UV.

KK reducing the NS5-brane theory on $S^3$, meaning at low energies compared to the inverse radius of the $S^3$, 
we obtain an ${\cal N}=1$ supersymmetric theory of a $U(N)$ gauge field, with a Chern-Simons term due to the $H$ field on $S^3$, 
coupled to gauginos, and all the rest of the fields are massive with $m\sim 1/R_{S^3}$. 
If the original level of the CS term in 6 dimensions is $k_6$, we can integrate out the massive gauginos, and obtain
a pure CS theory with level $k=k_6-N/2$. More precisely, since $U(N)=\frac{U(1)\times SU(N)}{Z_N}$, we have a
$(U(1)_{k_6}\times SU(N)_k)/Z_N$ theory, with a Witten index 
\be
I=\frac{(N+k-N/2)!}{(k-N/2)!N!}\;,
\ee
so we have a single vacuum for $k=N/2$, and supersymmetry for $k>N/2$, and susy breaking if $k<N/2$. This is then a very simple 
theory.

However, it turns out that sometimes one needs to consider the full 6-dimensional theory, as we will see.

\section{Penrose limit of Maldacena-Nastase solution}

\subsection{PP wave}

The Penrose limit corresponds to going near the null geodesic in the gravity dual, and can be understood (as Penrose showed) by 
considering the expansion in a parameter $R$ characterizing the background, and standard rescalings of the coordinates (the null 
coordinate transverse to the geodesic by $R^2$ and the other coordinates by $R$, keeping the direction of the geodesic unrescaled)
of the Penrose form of the metric, after which we obtain the pp wave in Rosen coordinates. 

Then in holography, it is better to consider a null geodesic that moves along an isometry direction of the gravity dual, since this would 
correspond in the field theory to states with a large charge $J$ in the R-symmetry direction corresponding to it. We will however see
that we don't have this luxury here, although we will initially try. The general procedure for the Penrose limit was considered in 
\cite{Gimon:2002sf} (see also examples in  \cite{Itsios:2017nou}), and involves writing an effective Lagrangian for the motion 
in the possible directions, but here we will not need to use it, as we can directly find the pp wave in Brinkmann coordinates, like 
in the $AdS_5\times S^5$ case.

It is now not possible to consider pp waves for the full solution, among other things because it is only available numerically, and also 
because it would be too complicated to disentangle. But we can try to use the theory in the IR, and correspondingly we should map 
to the field theory in the IR. The D5-brane string frame metric, that will be used for comparison with the field theory (as is usual) is 
\be
ds^2_{\rm 10, D, string}=e^{\Phi_D(\rho)}\left\{d\vec{x}_{2,1}^2+\a' N\left[d\rho^2+R^2(\rho)d\Omega_3^2
+\frac{1}{4}\left(\tilde w_L^a-A^a\right)^2\right]\right\}\;,
\ee
and we have $R^2\simeq \rho^2$, 
$\Phi_D(\rho)=\Phi_{D,0}+b'\rho^2$, with $b'\equiv \frac{2+3b^2}{8}$. However, it turns out that, using the $A^a$ in (\ref{MNsol})
leads to problems with the calculation so, instead, we can use a gauge-transformed version (given that the solution has a residual, 
spacetime-independent, gauge invariance), that removes the constant term and changes $w_L^a$ to $w_R^a$:
\be
A^a(\rho)\simeq \left(1+\frac{b}{2}\rho^2\right)w_L^a\Rightarrow A'^a(\rho)\simeq \frac{b}{2}\rho^2w_R^a.
\ee

The metric near $\rho=0$ becomes
\be
ds^2_{\rm 10, D, string}\simeq e^{\Phi_{D,0}+b'\rho^2}\left\{d\vec{x}_{2,1}^2+\a' N\left[d\rho^2+\rho^2d\Omega_3^2
+\frac{1}{4}\left(\tilde w_L^a-\frac{b}{2}\rho^2 w_R^a\right)^2\right]\right\}.
\ee

The geodesic we are interested in will be at $\rho=0$, as we already mentioned that we are interested in the IR of the gravity dual, 
it will be on an equator of the sphere at infinity $S_\infty^3$ defined by $\tilde\theta=0$ and $\tilde\psi=\tilde\phi$ (or 
$\tilde \phi_-\equiv (\tilde \phi-\tilde \psi)/2=0$), and it will move along $\tilde \phi_+\equiv (\tilde \phi+\tilde\psi)/2$, but we will also need a 
component of the motion along the internal sphere $S^3$, along $\phi$, in order to obtain a geodesic, and obtain the pp wave 
in Brinkmann coordinates.

We see that the relevant scale in the metric is $R^2=\a' N e^{\Phi_{D,0}}$, which should be taken to infinity for the Penrose limit, 
but it is convenient to introduce an auxiliary dimensionless number $L\rightarrow \infty$, such that 
\be
\mu^2=\frac{L^2}{\a' N e^{\Phi_{D,0}}}={\rm fixed}\;,\;\; R=\frac{L}{\mu}\rightarrow\infty.
\ee

The reason to introduce this factor is so that we can rescale with it the field theory coordinates, as
\be
e^{\frac{\Phi_{D,0}}{2}}dx^\mu=Ldx'^\mu\;,
\ee
though in the following, we will drop the primes on $x'^\mu$.

Further, we change variables from the angles $(\phi,\tilde\phi,\tilde\psi)$ to $(\hat \phi,\tilde\phi+,\tilde \phi_-)$, 
\bea
\tilde\phi_+=\frac{\tilde\phi+\tilde\psi}{2}\;, &
\tilde\phi_-=\tilde\phi-\tilde\phi_+=\frac{\tilde \phi-\tilde\psi}{2}\;, &
\hat\phi=\phi-\frac{b}{2}\tilde\phi_+=\phi-\frac{b}{2}\frac{\tilde\phi+\tilde\psi}{2}\Rightarrow\cr
\phi=\hat\phi+\frac{b}{2}\tilde\phi_+\;,&
\tilde\phi=\tilde\phi_-+\tilde\phi_+\;,&
\tilde\psi=\tilde\phi_+-\tilde\phi_-.
\eea

Next, we take the Penrose limit for motion on $\tilde\phi_+$, so that we rescale by $R=L/\mu$ as
\be
\rho=\frac{\mu}{L}r\;,\;\;
\tilde\theta=\frac{\mu}{L} 2v\;,\;\;
x^+=t\;,\;\;
x^-=L^2\left(t-\frac{\tilde\phi_+}{\mu}\right)\Rightarrow 
\tilde\phi_+=\mu x^+-\mu\frac{x^-}{L^2}.
\ee

All in all, we have then 
\be
\phi=\hat \phi+\frac{b}{2}\mu x^+ -\frac{\mu}{2L^2} b x^-\;,\;\;
\tilde \phi=\tilde \phi_-+\mu x^+-\frac{2\mu}{L^2}x^-\;, t=x^+\;,\;\;
\tilde \psi=\tilde\phi_+-\tilde\phi_-\;,
\ee
and using that, in the above Penrose limit we have 
\be
w_R^a\tilde w_L^a=2(d\phi d\tilde\phi_++\cos\theta d\psi d\tilde\phi_+)\;,
\ee
(but note that away from it, there are terms with $\cos\tilde\psi,\sin\tilde \psi$!), we obtain 
\bea
ds^2_{\rm 10, D, string}&\simeq & -\left(L^2-b' \mu^2 r^2\right)dt^2+ \sum_{i=1,2}dx_idx_i +dr^2+r^2(d\theta^2+d\phi^2+d\psi^2+
2\cos\theta d\phi d\psi)\cr
&&+dv^2+v^2d\tilde\phi^2+\left(\frac{L^2}{\mu^2}-b'r^2\right)d\tilde \phi_+ -2v^2d\tilde\phi d\tilde\phi_+
-br^2(d\phi d\tilde\phi_++\cos\theta d\psi d\tilde\phi_+)\cr
&=&-2dx^+dx^--\mu^2\left(\frac{b^2}{4}r^2+v^2\right)(dx^+)^2+\sum_{i=1,2}dx_idx_i+dv^2+v^2 d\tilde \phi_-^2\cr
&&+dr^2+r^2(d\theta^2+d\psi^2+d\hat \phi^2+2\cos\theta d\psi d\hat \phi).
\eea

Further defining 
\bea
dv^2+v^2d\tilde\phi_-^2&\equiv & dv_1^2+dv_2^2\cr
dr^2+r^2(d\theta^2+d\psi^2+d\hat \phi^2+2\cos\theta d\psi d\hat \phi)&\equiv & \sum_{i=1}^4 du_i^2\;,
\eea
we obtain the Brinkmann form of the pp wave, 
\be
ds^2=-2dx^+dx^--\mu^2\left(\frac{b^2}{4}\sum_{i=1}^4 u_i^2+v_1^2+v_2^2\right)(dx^+)^2+\sum_{i=1}^2 dx_i^2 +dv_1^2+dv_2^2
+\sum_{i=1}^4 du_i^2.
\ee

From our coordinate changes, we also obtain 
\bea
H&=&-p_+=i\d_+=i\d_t-\mu\left(\frac{b}{2} i\d_\phi+i\d_{\tilde\phi}+i\d_{\tilde\psi}\right)\cr
&=&E-\mu \left(\frac{b}{2}J_\phi+J_{\tilde\phi}+J_{\tilde\psi}\right)\equiv E-\mu J\cr
P^+&=&-\frac{1}{2}p_-=\frac{i}{2}\d_-=\frac{\mu}{L^2}\left(\frac{b}{2}J_\phi+J_{\tilde\phi}+J_{\tilde\psi}\right)\equiv \frac{\mu}{L^2}J.
\eea

We need to take the Penrose limit on the B field next. In the Penrose limit, since $w(\rho)\simeq 1-b\mu^2 r^2/L^2$, so 
$w^3(\rho)-3w(\rho)+2\simeq 0+{\cal O}(1/L^2)$, $h=0$, and since $w_L^1,w_L^2\sim 1/L$, $w_L^3\sim 1$, 
$A^a\simeq b\rho^2 w_R^a/2\sim w_R^a/L^2$, $F=dA+A\wedge A\sim dA
\sim w_R^a/L^2$, the only terms surviving in the D5-brane $H^{(D)}=dB^{(D)}$ are 
\be
H^{(D)}\simeq \frac{e^{\Phi_{D,0}}N}{4}\left[-\tilde w_L^1\wedge \tilde w_L^2\wedge \tilde w_L^3-dA^3\wedge \tilde w_L^3\right]\;,
\ee
and (since $\mu^2/L^2=(\a' N e^{\Phi_{D,0}})^{-1}$)
\bea
dA^a&=&b\rho d\rho\wedge w_R^a-\frac{b\rho^2}{2}dw_R^a\Rightarrow\cr
dA^3\wedge \tilde w_L^3&=&\frac{2\mu}{\a' N e^{\Phi_{D,0}}}\left(brdr\wedge w_L^3-\frac{br^2}{2}dw_R^3\right)\wedge dx^+\;,\cr
\tilde w_L^1\wedge \tilde w_L^2\wedge\tilde w_L^3&=&\sin\tilde\theta d\tilde\theta\wedge d\tilde\phi\wedge d\tilde\psi\cr
&\rightarrow& \frac{8\mu^3}{L^2}vdv\wedge d\tilde \phi_-\wedge dx^+=\frac{8\mu^3}{L^2}dv_1\wedge dv_2\wedge dx^+\;,
\eea
we obtain 
\be
H^{(D)}=\frac{\mu}{\a'}\left[\frac{b}{2}\left(rdr\wedge w_R^3-\frac{r^2}{2}dw_R^3\right)-2dv_1\wedge dv_2\right]\wedge dx^+.
\ee

Further, using the $u_i$ Cartesian coordinates, where 
\bea
u_1=r\cos\frac{\theta}{2}\cos\frac{\phi+\psi}{2}\;,&& u_2=r\cos\frac{\theta}{2}\sin\frac{\phi+\psi}{2}\;,\cr
u_3=-r\sin\frac{\theta}{2}\sin\frac{\phi-\psi}{2}\;, && u_4=r\sin\frac{\theta}{2}\cos\frac{\phi-\psi}{2}\;,
\eea
so that $rdr\wedge w_R^3-\frac{r^2}{2}dw_R^3=2[du_1\wedge du_2+du_3\wedge du_4]$, we finally find 
\be
H^{(D)}=dB^{(D)} =-2\frac{\mu}{\a'} \; dx^+\wedge \left[dv_1\wedge dv_2-\frac{b}{2}(du_1\wedge du_2+du_3\wedge du_4)\right]\;,
\ee
so that we can choose a gauge where we have 
\be
B^{(D)}=-2\frac{\mu}{\a'} x^+\left[dv_1\wedge dv_2-\frac{b}{2}(du_1\wedge du_2+du_3\wedge du_4)\right].
\ee

\subsection{String quantization and eigenstates}

Choosing the usual light-cone gauge $x^+=\tau$ and unit gauge $\gamma_{ab}=\eta_{ab}$, thus imposing as usual that the length of the 
string is $l=2\pi \a' p^+$, the string action in the pp wave is 
\bea
S&=&-\frac{1}{4\pi \a'}\int d\tau\int_0^{2\pi \a' p^+} 
d\sigma\left[\eta^{ab}\left(\sum_{i=1}^2 \d_a x_i\d_b x_i+\sum_{i=1}^2\d_a v_i\d_b v_i+\sum_{i=1}^4
\d_a u_i\d_b u_i\right)\right.\cr
&&\left.-\mu^2\left(\frac{b^2}{4}\sum_{i=1}^4 u_i^2+\sum_{i=1}^2v_i^2\right)\right]\cr
&&+\frac{2\mu}{4\pi \a'}\int d\tau\int_0^{2\pi \a' p^+} 
d\sigma\; \epsilon ^{ab}\tau\left[\d_a v_1\d_b v_2-\frac{b}{2}(\d_a u_1\d_b u_2+\d_a u_3\d_b u_4)\right].
\eea

We see that $x_1$ and $x_2$ are massless and non-interacting. 
The equations of motion of the fields $v_1,v_2,u_1,u_2,u_3,u_4$ are 
\bea
(-\d_\tau^2+\d_\sigma^2-\mu^2) v_1-2\mu  \d_\sigma v_2=0\;,&&
(-\d_\tau^2+\d_\sigma^2-\mu^2) v_2+2\mu  \d_\sigma v_1=0\cr
\left(-\d_\tau^2+\d_\sigma^2-\frac{b^2}{4}\mu^2\right)u_1-\mu  b \d_\sigma u_2=0\;, &&
\left(-\d_\tau^2+\d_\sigma^2-\frac{b^2}{4}\mu^2\right)u_2+\mu  b \d_\sigma u_1=0\cr
\left(-\d_\tau^2+\d_\sigma^2-\frac{b^2}{4}\mu^2\right)u_3-\mu  b \d_\sigma u_4=0\;,
&&\left(-\d_\tau^2+\d_\sigma^2-\frac{b^2}{4}\mu^2\right)u_4+\mu  b \d_\sigma u_3=0.\cr
&&
\eea

Next we choose free waves (Fourier modes), 
\be
v_i=v_{i,0}\exp[-i\omega \tau+ik_i\sigma]\;,
\ee
and similarly for the others, and with the usual rescaling by $p^+$ of the gauge condition as above, we have for the quantization of 
momenta around the $\sigma$ circle
\be
k_{i,n}=\frac{n_i}{\a' p^+}\;,
\ee
so that for the $v_i$'s we obtain 
\be
(\omega^2-k_1^2-\mu^2)v_{1,0}-2i\mu k_2 v_{2,0}=0\;,\;\;
+2i\mu k_1 v_{1,0}+(\omega^2-k_2^2-\mu^2)v_{2,0}=0.
\ee

From the vanishing of the determinant, we obtain the eigenvalues
\bea
\omega^2_{v,\pm}&=&\frac{1}{2}\left[2\mu^2+k_1^2+k_2^2\pm \sqrt{(k_1^2-k_2^2)^2+16\mu^2 k_1k_2}\right]\Rightarrow\cr
\omega_{v,\pm}&=&\mu\sqrt{1+\frac{n_1^2+n_2^2}{2(\mu\a'p^+)^2}\pm \sqrt{\frac{(n_1^2-n_2^2)^2}{4(\mu\a'p^+)^4}+4\frac{n_1n_2}{
(\mu\a'p^+)^2}}}.
\eea

Similarly, for the $u_i$ modes we obtain (two modes for each)
\bea
\omega^2_{u,\pm}&=&\frac{1}{2}\left[\frac{b^2\mu^2}{2}+k_1^2+k_2^2\pm \sqrt{(k_1^2-k_2^2)^2+4b^2 \mu^2 k_1k_2}\right]\Rightarrow\cr
\omega_{u,\pm}&=&\mu\sqrt{\frac{b^2}{4}+\frac{n_1^2+n_2^2}{2(\mu \a' p^+)^2}\pm \sqrt{\frac{(n_1^2-n_2^2)^2}{4(\mu \a' p^+)^4}
+b^2\frac{n_1n_2}{(\mu\a'p^+)^2}}}.
\eea

We see then that, for $n_i=0$ or $\mu \a' p^+\rightarrow\infty$, we have $\omega_{v,\pm}/\mu=1$ and $\omega_{u,\pm}/\mu=b/2$, which 
at the supersymmetric point equals 1/6.

The $x_1,x_2$ modes are massless and non-interacting, so for them we have 
\be
\omega_{x_i}=\mu \frac{n_i}{(\mu \a' p^+)}.
\ee

Moreover, we see that for $n_1=n_2\equiv n$, the $v$ eigenvalues become
\be
\omega_{v,\pm}=\sqrt{1+\frac{n^2}{(\mu \a' p^+)^2}\pm \frac{2n}{\mu \a'  p^+}}=1\pm \frac{n}{\mu \a' p^+}\;,\label{result}
\ee
and the $u$ eigenvalues become
\be
\omega_{u,\pm}=\sqrt{\frac{b^2}{4}+\frac{n^2}{(\mu \a' p^+)^2}\pm \frac{2b/2}{\mu \a' p^+}}=\frac{b}{2}\pm \frac{n}{\mu \a' p^+}.
\ee

We see that, except for the different leading term ($0,1$ or $b/2$), the first correction is universal ($n/(\mu \a' p^+)$) and exact. 
This is what we will attempt to reproduce from the field theory side.

\section{Large charge sector and spin chain}

We now focus on the field theory side, and see if we can again understand the Penrose limit in terms of a spin chain.

\subsection{Field theory modes}

As we have reviewed in section 2, the field theory on the 5-branes is ${\cal N}=(1,1)$ supersymmetric in 5+1 dimensions, and it KK reduces 
on $S^3$ with the twist to a 2+1 dimensional ${\cal N}=1$ supersymmetric gauge theory, made up of 
pure Chern-Simons gauge fields, with the gauginos that can be integrated out, and this is coupled to the KK modes. 

It will turn out that the KK modes also must be taken into account, so we consider the 6-dimensional theory, and its expansion on $S^3$. 
In 6 dimensions we have the ${\cal N}=1$ vector multiplet (with one vector and one Weyl spinor), and an ${\cal N}=1$ hypermultiplet 
(which contains 2 complex scalars and one Weyl spinor). The 2 complex scalar (or 4 real) correspond to the 4 coordinates transverse to the 
5-brane, and transform under the R-symmetry $SO(4)\simeq SU(2)_L\times SU(2)_R$. In order to keep ${\cal N}=1$ susy after 
compactifying on $S^3$, we must embed the $SU(2)_{\rm tg.}$-valued spin connection on $S^3$ into the R-symmetry, specifically 
into $SU(2)_L$, which means that under KK reduction we gauge $SU(2)_L$ with the spin connection.

The 6-dimensional theory with one gauge field, 2 Weyl spinors and 2 complex scalars decomposes, after the KK reduction but before the 
twist, under the remaining symmetry group $SO(2,1)\times SU(2)_{\rm tg.}\times SU(2)_L\times SU(2)_R$ (note the decomposition 
$SO(5,1)\rightarrow SO(2,1)\times SU(2)_{\rm tg.}$), as 
\be
{\rm gauge}: \left(\mathbf{3},\mathbf{1},\mathbf{1},\mathbf{1}\right)\oplus\left(\mathbf{1},\mathbf{3},\mathbf{1},\mathbf{1}\right)\;,\;\;
{\rm scalars}: \left(\mathbf{1},\mathbf{1},\mathbf{2},\mathbf{2}\right)\;,\;\;
{\rm fermions}:  \left(\mathbf{2},\mathbf{2},\mathbf{2},\mathbf{1}\right)\oplus\left(\mathbf{2},\mathbf{2},\mathbf{1},\mathbf{2}\right).
\ee

After we twist by embedding the $SU(2)_{\rm tg.}$ spin connection into $SU(2)_L$, we have 
$SO(2,1)\times \left(SU(2)_{\rm tg.}\times SU(2)_L\right)_{\rm diag.}\times SU(2)_R$, so we obtain 
\be
{\rm gauge}: \left(\mathbf{3},\mathbf{1},\mathbf{1}\right)\oplus\left(\mathbf{1},\mathbf{3},\mathbf{1}\right)\;,\;\;
{\rm scalars}: \left(\mathbf{1},\mathbf{2},\mathbf{2}\right)\;,\;\;
{\rm fermions}: \left(\mathbf{2},\mathbf{1},\mathbf{1}\right)\oplus\left(\mathbf{2},\mathbf{3},\mathbf{1}\right)
\oplus\left(\mathbf{2},\mathbf{2},\mathbf{2}\right)\;,
\ee
which gives the field content of the ${\cal N}=1$ 3-dimensional theory. Of course, now we need to find the masses of these fields
(and remember that massless fermions are integrated out, modifying the level $k$ of the Chern-Simons theory). 
The massless fields are the singlets under the spin connection, so $(SU(2)_{\rm tg.}\times SU(2)_L)_{\rm diag}$, so only a 
gauge field and a fermion.

We are mostly interested in the bosonic degrees of freedom so, besides the massless gauge fields, we have the massive 
scalars in the $\left(\mathbf{1},\mathbf{3},\mathbf{1}\right)$ (from the KK expansion of the 6-dimensional gauge fields) and 
$\left(\mathbf{1},\mathbf{2},\mathbf{2}\right)$ (from the KK expansion of the 6-dimensonal scalars) representations. 
We would like to calculate their 3-dimensional masses and representations for the symmetry groups, and see which ones can be
used to compose operators dual to string modes. 

We KK expand on the sphere, written as a coset manifold, $S^3=SO(4)/SO(3)\equiv G/H$, or more precisely, with $G=SO(4)\simeq SU(2)'_L
\times SU(2)'_R$ and $H=SU(2)'_{\rm diag.}$ (the diagonal part of the two $SU(2)$'s) though, since $SU(2)'_{\rm diag}=SU(2)_{\rm tg.}$ 
is also the tangent space symmetry of the spin connection, via the twisting, it becomes also diagonal with respect to the 
$SU(2)_L$ R-symmetry.

In general, for fields that are scalars after dimensional reduction (here, $SO(2,1)$ scalars), the KK expansion takes the form
\be
\phi_M(x,y)=\sum_{q, I_q}\phi^{q,I_q}(x) Y_M^{q,I_q}(y)\;,
\ee
where $I_q$ is an index in a representation of the isometry group $G$ of the coset space, here $SO(4)=SU(2)'_L\times SU(2)'_R$, with 
dimension $q$, and $M$ is an index in the local Lorentz group $H$ of the coset space, here $SO(3)=SU(2)_{\rm diag.}$, and 
$Y_M^{q,I_q}(y)$ is the spherical harmonic. In general, $M$ can also include the indices in the internal group, unrelated to $G$ or 
$H$, in this case the $SU(2)_R$ R-symmetry.
Since representations of $SU(2)$ are defined by a spin $j$ and have index $m=-j,...,+j$, the representations $q$ of $G$ are defined by 
$(j_1,j_2)$, with index $I_q=(m_1,m_2)$. We can therefore write for the scalars, in general
\be
\phi_{m_1'(m'_2)}=\sum_{j_1,j_2}\sum_{m_1,m_2} \phi^{(j_1,j_2)}_{m_1m_2}(x)Y^{(j_1,j_2); m_1m_2}_{m'_1(m'_2)}(y).
\ee

The 3-dimensional masses for these fields are obtained, as usual, by splitting the 6-dimensional Laplacian 
as $\Box_{x,y}=\Box_x+\Box_y$, with the spherical harmonics being eigenfunctions of $\Box_y$ on the $S^3$, with eigenvalue
\cite{Nicolai:2003ux,BenAchour:2015aah}
\be
\Box_y Y^{(j_1,j_2)m_1m_2}_{m'_1(m'_2)}(y)=-\frac{j_1(j_1+1)+j_2(j_2+1)+s(s+1)}{R^2}Y^{(j_1,j_2)m_1m_2}_{m'_1(m'_2)}(y)\;,
\ee
where $R$ is the radius of $S^3$, 
$j_1,j_2$ are dimensions of representations under $SU(2)'_L\times SU(2)'_R$, and $s$, or "spin", from the point of view of the 
$SO(3)=SU(2)'_{\rm diag}=(SU(2)'_L\times SU(2)'_R)_{\rm diag}$ is the representation of the diagonal group ("addition of angular momenta"),
so is from $|j_1-j_2|$ to $j_1+j_2$. 

For $\left(\mathbf{1},\mathbf{3},\mathbf{1}\right)$ we have $s=1$, meaning $j_1=j_2=1/2$ for the lightest fields, so for them
$m^2=7/(2R^2)$, whereas for $\left(\mathbf{1},\mathbf{2},\mathbf{2}\right)$ we have $s=1/2$, so $j_1=1/2, j_2=0$ for the lightest 
fields, so for them $m^2=3/(2R^2)$.

Although they have different masses, both of these lightest modes are needed to match to the string side, since they have different 
symmetry properties (and provenance).

In the 4-dimensional Maldacena-N\'{u}\~nez case, the resulting theory of massive scalars and 
fermions coupled to gauge fields was able to 
``deconstruct'', through a fuzzy $S^2$, the 6-dimensional theory, at least at the classical level, 
as shown by Andrews and Dorey \cite{Andrews:2006aw}. In our 3-dimensional case, this is considerably 
more difficult, since the construction of the fuzzy $S^3$ is much less clear, though the ingredients of
massive scalars and fermions are likely still crucial.

\subsubsection{Matching to string modes}

Corresponding to the strings in the pp wave, in their discretized form, we expect to find large charge operators in the field theory, 
just like we had seen in the case of BMN operators for ${\cal N}=4 $ SYM in \cite{Berenstein:2002jq}. The difference is that now 
the operators are in a confining theory, with a nontrivial IR dynamics. That is why these gauge-invariant
operators were interpreted in the 4-dimensional confining cases as creating some ring-like hadrons dubbed "annulons" in 
\cite{Gimon:2002nr}. Other than this, we expect the same kind of qualitative construction. 

To match to string modes, we remember that the $v_1,v_2$ modes were obtained from $\tilde\theta,\tilde\phi\in S^3_\infty$, 
the $u_1,...,u_4$ modes were obtained from $r$ and $S^3$ modes, and $x_1,x_2$ are the field theory space coordinates. 

On the field theory side, the three $\left(\mathbf{1},\mathbf{3},\mathbf{1}\right)$ modes $A_a$ are modes on the $S^3$, the 
four $\left(\mathbf{1},\mathbf{2},\mathbf{2}\right)$ modes $\phi^M$ transform under the R-symmetry corresponding to the 
$SO(4)=SU(2)_L\times SU(2)_R$ invariance of the transverse sphere $S^3_\infty$, and other relevant objects that can act 
as string oscillators in the pp wave are the covariant derivatives $D_t$ and $D_{x_i}$. 

It becomes clear then that $D_{x_i}$ (really, in the two sphere directions in the spherical coordinate split of the Euclidean version 
of the 3-dimensional field theory coordinate) should correspond to string oscillators in the $x_1,x_2$ directions, and the $\phi^M$ 
should be split according to the charge $J$ into some $Z$ and $\bar Z$ fields corresponding to $x^+$ and $x^-$
(for motion mostly on $S^3_\infty$, under a combination of $\tilde\phi$, $\tilde\psi$, and also $\phi$), and 
two fields $W$ and $\bar W$ corresponding to string oscillators in the two $v_1,v_2$ directions. 
Finally, the $A_a$ modes and the $D_t$ covariant derivative (really, in the radial Euclidean time direction for the Euclidean 
version of the 3-dimensional field theory coordinate) should correspond to the string oscillators in the $u_1,...,u_4$ directions. 

We confirm this by charge assignments and calculation of the corresponding energies in the free case (at zero coupling, or 
no excitations: the BPS case). We first decompose the $\phi^M$ modes using the bifundamental action of $SO(4)=SU(2)_L\times
SU(2)_R$, by $\phi^M\rightarrow V_L\phi^M V_R$, as (using $\sigma_M\equiv(i\one,\sigma_j)$, for $M=0,1,2,3$ and $j=1,2,3$)
\be
\Phi^{\a\b}=\frac{1}{\sqrt{2}}(\sigma_M)^{\a\b}\phi^M=\frac{1}{\sqrt{2}}\begin{pmatrix}i\Phi^0+\Phi^3 & \Phi^1-i\Phi^2\\
\Phi^i+i\Phi^2 & i\Phi^0-\Phi^3\end{pmatrix}=\frac{1}{\sqrt{2}}\begin{pmatrix}Z& W^*\\ W& -Z^*\end{pmatrix}.
\ee

The charges $J_{\tilde\phi}$ and $J_{\tilde\psi}$ correspond to {\em independent} $U(1)$ rotations inside $S^3_\infty$, 
which means that one of them is a $U(1)\subset SU(2)_L$, and the other a $U(1)\subset SU(2)_R$. But from the above decomposition, 
the charges of $Z$ under $U(1)\subset SU(2)_L$ and $U(1)\subset SU(2)_R$ are the same, and also the same as the charge of 
$W$ under $U(1)\subset SU(2)_L$, and the opposite as the charge of $W$ under $U(1)\subset SU(2)_R$. Of course, neither of the 
$\Phi^M$ is charged under an $U(1)\subset (SU(2)'_L\times SU(2)'_R)_{\rm diag.}$, that corresponds to $J_\phi$.
Also, $A_a$ {\em is} charged under $U(1)\subset (SU(2)'_L\times SU(2)'_R)_{\rm diag.}$, and because of the twisting, so is 
$D_t$ (corresponding to the scaling direction of the field theory, holographically identified with motion in $\rho$, and 
$\rho, S^3_\infty$ make up the transverse coordinates to the 5-brane), and with the same charge. 

The normalization of the $J_{\tilde \phi}$ and $J_{\tilde\psi}$ and of the $J_\phi$ can be chosen according to our needs. 
It will be useful to fix the basic $J_{\tilde \phi}$ and $J_{\tilde\psi}$ charges of $Z$ as $-1/4$, and the basic $J_\phi$ charge of $A_a$ 
as $1/2$. Another important point is that, in the Hamiltonian coming from the field theory modes, a total constant is irrelevant, 
so we will be subtracting a value of $1$ from each field. 

Then, remembering that $H=E- J$, $J=\frac{b}{2}J_\phi+J_{\tilde\phi}+J_{\tilde\psi}$ and $P^+=J/R^2$, 
that $E\rightarrow \Delta$ in field theory, and that, as we said, we can subtract an $E_0=1$ from each field, we obtain 
the table:
\begin{center}
    \begin{tabular}{|c|c|c|c|c|c|c|c|}
    \hline
    field & $Z$ & $W$ & $\bar Z$ & $\bar W$ & $A_a$ & $D_t$ & $D_{x_i}$ \\
    \hline\hline
    $\Delta$ & 1/2 & 1/2 & 1/2 & 1/2 & 1 & 1 & 1 \\
    \hline
    $J_\phi$ & 0 & 0 & 0& 0 & 1/2 & 1/2 & 0\\
    \hline
    $J_{\tilde\phi}$ & -1/4 & -1/4 & +1/4 & +1/4 & 0 & 0 & 0 \\
    \hline
    $J_{\tilde \psi}$ & -1/4 & +1/4 & +1/4 & -1/4 & 0 & 0 & 0\\
    \hline
    $J$ & -1/2 & 0 & +1/2 & 0 & b/4 & b/4 & 0 \\
    \hline
    $\Delta-J$ & 1 & 1/2 & 0 & 1/2 & 1-b/4 & 1-b/4 & 1\\
    \hline
    $H/\mu=\Delta-J-E_0$ & 0 & -1/2 & -1 & -1/2 & -b/4 & -b/4 & 0\\
    \hline
    oscillator & - & $v_1$ & - & $v_2$ & $u_1,u_2,u_3$ & $u_4$ & $x_1,x_2$\\
    \hline
    \end{tabular}
\end{center}

This is the correct value of the Hamiltonian, if we multiply by $\mu=-2$. Note that, as usual, $\bar Z$ doesn't give an 
oscillator ($Z$ and $\bar Z$ correspond to $X^+$ and $X^-$).

As usual then, in the standard (BMN) limit, with $J\rightarrow\infty $ and $H=\Delta-J-E_0$ fixed, we obtain the string vacuum 
$|0;p^+\rangle$ constructed out of $J$ fields $Z$, 
\be
|0;p^+\rangle=\frac{1}{\sqrt{J} N^{J/2}}\Tr[Z^J]\;,
\ee
and string oscillators $a^{\dagger Q}_n$, for $Q:(x_1,x_2,v_1,v_2,u_1,u_2,u_3,u_4)$ correspond to 
insertions of $\Phi^Q=(D_{x_1},D_{x_2},W,\bar W,A_1,A_2,A_3,D_t)$, giving the (BMN) operators as 
\be
a^{\dagger Q_1}_n a^{\dagger Q_2}_{-n}|0;p^+\rangle=\frac{1}{\sqrt{J}}\sum_{l=1}^J\frac{1}{N^{\frac{J}{2}+1}}\Tr[\Phi^{Q_1}Z^l
\Phi^{Q_2}Z^{J-l}]e^{\frac{2\pi i nl}{J}}.
\ee

The fermions could be treated similarly, but we will not consider them, as we have also done in the string case.

\subsection{Interactions and corrections to anomalous dimension}

Having obtained matching at the BPS ($n_i=0$) or free ($g^2=0$) level, we consider the interactions next. 

The 6-dimensional bosonic action for the (Lorentzian) D5-branes can be written as (it is a dimensional reduction of 10-dimensional 
${\cal N}=1$ SYM)
\be
S_{\rm 6d, SYM}=\frac{1}{g^2_{YM, 6d}}\int d^6x \Tr\left[-\frac{1}{2}(D_\Pi \Phi_M)^2
-\frac{1}{4}F_{\Pi\Sigma}^2-\frac{1}{32}[\Phi^{\a\gamma},{\Phi^\b}_\gamma]
[\Phi_{\b\delta},{\Phi^\delta}_\a]\right]\;,
\ee
where $D_\Pi=\d_\Pi-i[.,A_\Pi]$ and $\Pi=0,..,5$ and the $\a,\b$ indices are raised and lowered with $\epsilon_{\a\b}$ 
like $\epsilon_{\a\b}\Phi^{\b \gamma}={\Phi_\a}^\gamma$, and the indices in the adjoint of $SU(N)$ are implicit. The gauge field from the 10-dimensional reduction is $(A_\Pi,A_M\equiv \Phi_M)$.

As we can see, the interactions are formally the same as in ${\cal N}=4$ SYM, as commutator square of the scalars.

After compactification to 3 dimensions and twisting, $\Phi_M$ (decomposed as $\Phi^{\a\b}$, and that as $Z,W,\bar Z,\bar W$)
generates a massive KK tower, out of which we consider only the lightest. Also $A_a$ generates a massive tower, out of which 
we consider only the lightest, but we will not consider these modes here, as they are more difficult to deal with. 
Denote by $m$ the mass of this lightest $\Phi_M$ (so $Z,W,\bar Z,\bar W$) mode. 

Its Euclidean propagator is then ($\theta,\tau,\rho,\sigma$ are adjoint $SU(N)$ indices)
\be
P\left(x,z\right)\equiv\left\langle Z\left(x\right)_{\theta}^{\tau}\bar Z\left(z\right)_{\rho}^{\sigma}\right\rangle 
=-\delta_{\theta}^{\sigma}\delta_{\rho}^{\tau}\frac{g^2_{YM}}{4\pi}\frac{\exp\left[-m|x-z|\right]}{|x-z|}.\label{prop}
\ee

Here we have used the coupling of the KK reduced theory, 
\be
g^2_{YM}=\frac{g^2_{YM,6d}}{V_{S^3}}=\frac{(2\pi)^3\a' e^{\Phi_D,0}}{2\pi^2 R^3_{S^3}}=\frac{4\pi e^{\Phi_{D,0}}\a'}{r^3}\;,
\ee
where we have used that $g^2_{Dp}=(2\pi)^{p-2}e^{\Phi_0}\a'^{\frac{p-3}{2}}$ and $V_{S_3}=2\pi^2 R_{S^3}^3$, and 
$R_{S^3}=$ \\$e^{\Phi_{D,0}/2}\sqrt{\a' N}\rho=r$.

The relevant vertices, for interactions of $W$'s with $Z$'s are again of the type $[W,Z][\bar W,\bar Z]$ as in the 
$SU(2)$ sector of the ${\cal N}=4$ SYM in 4 dimensions, the only difference being the dimension of the fields, and 
the coupling. 

\begin{figure}[t!]
\begin{center}
\includegraphics[scale=0.3]{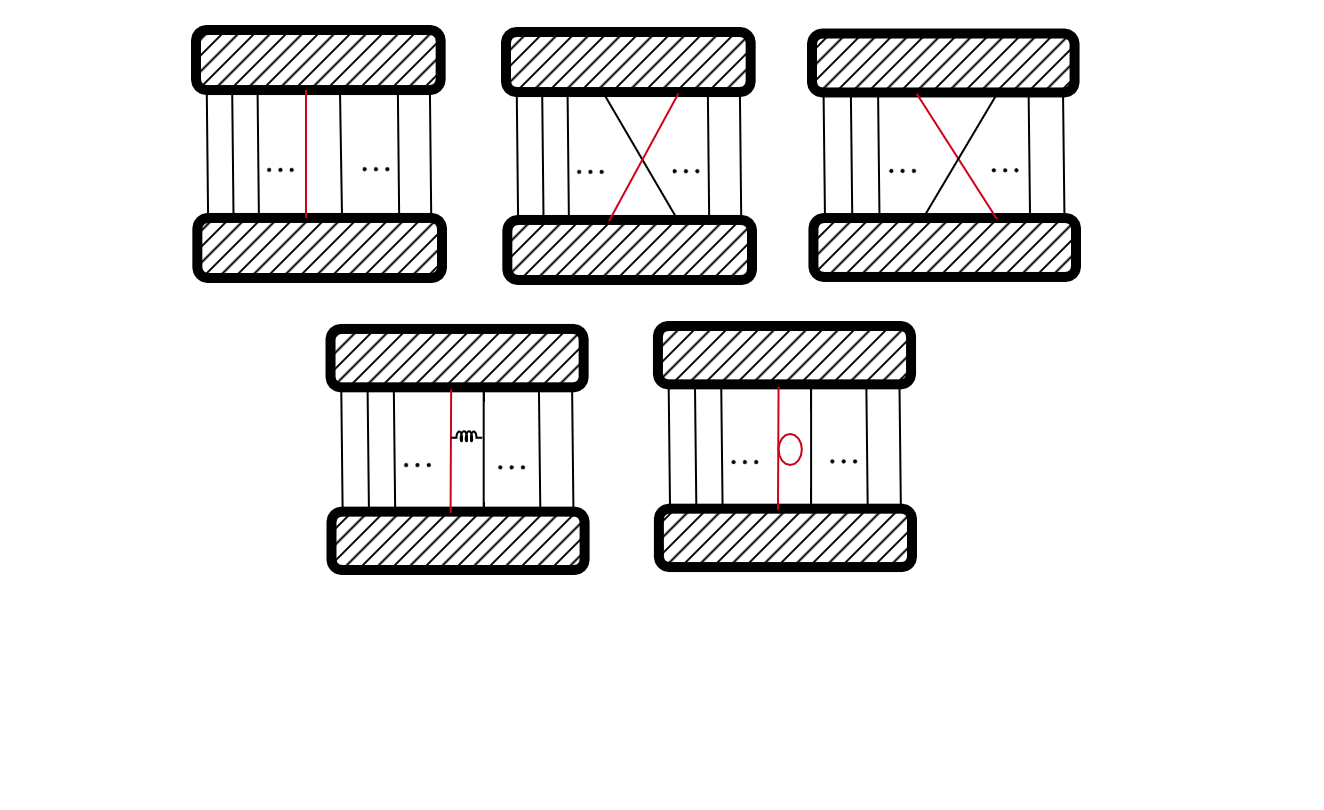}
\vspace{-2cm}
\caption{Diagrams for the interactions.}\label{pastedfigure}
\end{center}
\end{figure}

Corrections to the conformal dimension (and thus to the energy $H$) of the BMN operators ${\cal O}_\a$ 
with insertions of $W$ and $\bar W$ are 
obtained as usual from their 2-point functions \cite{Berenstein:2002jq}, which is given in terms of 
hopping terms for $W$ (or $\bar W$) on the chain of $Z$'s, 
giving a factor of $\left(\exp\left[-i\frac{2\pi n}{J}\right]+\exp\left[i\frac{2\pi n}{J}\right]\right)$, 
plus non-hopping terms that ensure that the BPS operators are not corrected, so giving the needed $-2$. 
The relevant diagrams are given in Fig.\ref{pastedfigure}.

The one-loop anomalous dimension correction to the 2-point function of BMN operators is then 
\bea
\left\langle \mathcal{O}\left(x\right)\mathcal{O}\left(0\right)\right\rangle &=&\left\langle \mathcal{O}\left(x\right)
\mathcal{O}\left(0\right)\right\rangle _{tree}\left[1+\right.\cr
&&\left.+\frac{N}{2g^2_{YM}}\left(\cos\left(\frac{2\pi n}{J}\right)-1\right)
\int d^{3}z\frac{\left[P\left(z,0\right)\right]^{2}\left[P\left(x,z\right)\right]^{2}}{\left[P\left(x,0\right)\right]^{2}}\right]\cr
&=&\left\langle \mathcal{\bar{O}}\left(x\right)\mathcal{O}\left(0\right)\right\rangle _{tree}\left[1
+\frac{g^2_{YM}N|x|}{2\left(4\pi\right)^{2}}I\left(x\right)\sin^{2}\left(\frac{n\pi}{J}\right)\right]\;,
\eea
where we have defined the dimensionless integral
\begin{equation}
I\left(x\right)=\frac{|x|}{\exp\left[-2m|x|\right]}\int d^{3}z\frac{\exp\left[-2m|z|\right]\exp\left[-2m|x-z|\right]}{|z|^{2}|x-z|^{2}}\label{eq:.}
\end{equation}
and we have 
\be
\left\langle \bar{\mathcal{O}}(x)\mathcal{\mathcal{O}}(0)\right\rangle _{tree}=\left(P\left(x,0\right)\right)^{J+2}.
\ee

If $n$ is fixed and $J\rightarrow \infty$, we get
\be
\left\langle \mathcal{O}\left(x\right)\mathcal{O}\left(0\right)\right\rangle =
\left(P\left(x,0\right)\right)^{J+2}\left[1+\frac{g^2_{YM}N|x|}{32J^2}n^2I(x)\right].
\ee

Note the appearance of the $x$-space effective dimensionless coupling $g^2_{YM}N|x|$ ($g^2_{YM}$ has mass dimension 1) in front
of the one-loop correction, as expected. More precisely, we obtain $g^2_{YM}N|x|/J^2$, which is to be matched with the pp wave 
quantity $1/(\mu \a' p^+)^2$, similar to the 4-dimensional ${\cal N}=4$ SYM case.

Contrary to the ${\cal N}=4$ SYM case in 4 dimensions, however, or to the 3-dimensional ABJM case reviewed in the next 
subsection, this integral is now both UV and IR convergent (it is in fact UV and IR convergent even if $m=0$), 
so we don't expect any $\log(|x|\Lambda)$ to appear, so 
it would formally seem like there is no correction to the anomalous dimension (since that comes out of expanding $|x|^{\Delta_0
+\delta \Delta}$ into $|x|^{\Delta_0}\left(1+\delta \Delta \ln (|x|\Lambda) \right)$).

However, note that $g^2_{YM}$ will renormalize (in an asymptotically free theory like the theory of massive scalars coupled 
to gauge fields that we describe here, with effective coupling $g^2_{YM}N|x|$) 
to $g^2_{YM}(\Lambda)$, which to a first approximation can be understood as $g^2_{YM}\log (|x|\Lambda)$, so in effect we have
the whole correction term in the square bracket being the correction to the anomalous dimension, so for $W$ and $\bar W$ we have
\be
\Delta-J-E_0=1+\frac{g^2_{YM}N|x|n^2}{32J^2}I(x).
\ee

But, as we said, we are interested in the UV limit, $|x|\rightarrow 0$, and $m$ is fixed, in which case
\be
I(x)\rightarrow |x|\int d^3x \frac{1}{|z|^2|x-z|^2}\sim |x|4\pi \int_{\Lambda'=|x|}^\infty z^2dz \frac{1}{|z|^4} =|x|4\pi \frac{1}{|x|}=4\pi\;,
\ee
so we obtain 
\be
\Delta-J-E_0=1+\frac{\pi}{8}\frac{g^2_{YM}N|x|n^2}{J^2}.
\ee

Moreover, the same calculation of the correction to $\Delta$ will also apply to the other "string oscillators", i.e., to insertions of 
$D_{x_i}$, $D_t$ and $A_a$, since in fact the D5-brane action reduces to the D3-brane action of ${\cal N}=4$ SYM (and both 
are reductions of the ${\cal N}=1$ SYM action in 10 dimensions), so the interactions are as democratic (equal) here as they were 
in the latter case, where they all give the same result. The only difference is, of course, the leading term, which is $0$ for $D_{x_i}$ 
and $b/2$ for $A_a$ and $D_t$. 

For comparison with the pp wave result, we note that, given that we rescaled $e^{\Phi_{D,0}/2}dx^\mu=Ldx'^\mu$, 
or $dx^\mu=\mu \sqrt{\a'} \sqrt{N} dx'^\mu$, $p^+=J/L^2$ needs rescaling by this factor, so 
\be
\frac{n^2}{(\a' \mu p^+)^2}=\frac{1}{(\mu\sqrt{\a'} \sqrt{N})^2}\frac{L^4n^2}{\mu^2\a'^2 J^2}=\frac{e^{2\Phi_{D,0}}Nn^2}{J^2}
=\frac{g^4_{YM}r^6Nn^2}{16\pi^2\a'^2 J^2}.
\ee

That means that, if we make the replacement 
\be
g^4_{YM}r^6\rightarrow g^2_{YM}4\pi^3 |x|\a'^2\;,
\ee
we obtain the subleading term with $n^2$ inside the square root in (\ref{result}), in the absence of the $\pm $ term that indicates 
the coupling of the two $v_i$ modes. 

We see then that we don't quite get the correct behaviour, most importantly we don't get the mixing of the two $v_i$ modes
(and, of course, we have considered $n_1=n_2$ already), and we don't even obtain the correct power of $g^2_{YM}$. 

There are several reasons why this can happen. The most important one is that, as we mentioned, 
we must consider the UV of the field theory, 
in order to have perturbation theory in the effective coupling $g^2_{YM}N|x|$. Of course, like in the case of ${\cal N}=4$ SYM, 
we see that we can still have $g^2_{YM}N|x|$ large if $g^2_{YM}N|x|/J^2$ is fixed, but now we are also forced to consider explicitly 
$|x|\rightarrow 0$. That is contrary to what we did in the pp wave case, where we considered the IR of the gravity dual. 

The second reason is that, as we saw, $J$ contained a piece from $J_\phi$, besides the $J_{\tilde\psi}$ and $J_{\tilde\phi}$. 
While $\tilde \psi$ and $\tilde\phi$ were isometries of the metric ($d\tilde\Omega_3^2=\tilde w_L^a\tilde w_L^a$ contains only 
$\cos\tilde\theta$), $\phi$ is not an isometry of the metric, except in the strict Penrose limit: $w_L^aw_L^a$ and $w_R^a w_R^a$ contain 
$\cos\theta$ only, but $\tilde w_L^aw_R^a$ (the cross term from $(\tilde w_L^a-\frac{b}{2}\rho^2 w_R)^2$) contains only $\cos\theta$ only 
in the strict Penrose limit, away from it, it contains also terms depending on $\phi$.
The fact that the null geodesic was not completely in an isometry direction was also a factor in \cite{Itsios:2017nou}, where also it 
was found that there was a mismatch at the level of the first correction to $\Delta-J$. 

The third  reason is that in general, once we have a smaller amount of supersymmetry, we expect that the 
field theory formula for $\Delta-J$ is modified, by $\sin^2(p/2)\simeq \pi^2 n^2/J^2$  
being multiplied by a function of the 't Hooft coupling, which takes one 
value at $\lambda=0$ (perturbation theory) and another at $\lambda=\infty$ (in the gravity dual). 
This is what happens in the case of the ABJM model, as we will review in the next subsection. In this case, moreover, we don't have a 
conformal theory, so the $g^2_{YM}$, $N$ and $|x|$ dependences are all independent.

Finally, and related to the previous reason, we have neglected the effect of other Feynman diagrams, 
in particular the ones with $\bar Z$ that, like in 
the case of ${\cal N}=4$ SYM, were assumed to vanish, as the field $\bar Z$ gets an infinite mass. But it is not clear that this reasoning 
is still valid in this case, of a confining theory with scale dependence of the coupling. 

Of course, there is also the possibility that there are further modes and couplings coming from the KK
expansion that need to be considered, beyond what we did in the previous subsection.

\subsection{Comparison with the ABJM and GJV models, and holographic cosmology cases}

Finally, in order to gain a clearer picture, let us consider other cases of three-dimensional spin chains obtained from gravity 
dual pairs. 

{\bf The ABJM model}

First off, the standard model in three dimensions, the ABJM model \cite{Aharony:2008ug}, is a conformal and ${\cal N}=6$ supersymmetric
$U(N)\times U(N)$ Chern-Simons gauge theory at level $(k,-k)$. 
Like in our case, the gauge fields are Chern-Simons type, but unlike in our case, 
we have bifundamental matter for the $SU(N)\times SU(N)$ gauge group, and conformality means, in particular, that the couplings 
to not run: in fact, the 't Hooft coupling in this case is discrete, $N/k$, with $k$ the Chern-Simons level. 

The Penrose limit and its spin chain dual in the field theory was analyzed by \cite{Nishioka:2008gz} (see also \cite{Cardona:2014ora}
for open strings on D-branes in this case)
The 4 complex scalar fields (for 4 chiral multiplets) are denoted by $(A_1,A_2,\bar B_1,\bar B_s)$, with $A_1,A_2$ in the 
$(N,\bar N)$ representation, and $B_1,B_2$ in the $(\bar N,N)$ representation of the gauge group.

In this case, the scalars are uniquely identified with coordinates in spacetime, and therefore their relation to the spacetime charges $J$
is well-defined. In field theory, the charges are for a combination of $U(1)$'s inside the $SU(4)$ R-symmetry. 
One obtains that $J(A_1)=J(B_1)=1/2$ and $J(A_2)=J(B_2)=0$, while $\Delta[A_1,A_2,B_1,B_2]=1/2$. However, in order to multiply 
some objects inside a trace, in order to construct spin chains, they need to transform under a single gauge group, which 
uniquely identifies the object from which the vacuum is constructed as the bilinears $A_1B_1$, so $|0;p^+\rangle\sim \Tr[(A_1B_1)^J$. 

The bosonic string oscillators inserted inside the trace are: $A_1B_2, A_1\bar A_2, A_2B_1, \bar B_2 B_1$ with $\Delta-J=1/2$ and 
$D_\mu $ ($\mu=0,1,2$) and $A_1\bar A_1+\bar B_2 B_2$ with $\Delta-J=1$. The spin chain completely reproduces
the pp wave result, though now (because we have less than maximal supersymmetry), the 
result is $\Delta-J=\frac{1}{2}\sqrt{1+16f^2(\lambda)\sin^2 p/2}$, where 
$f(\lambda\rightarrow 0)\simeq \lambda$ differs from $f(\lambda\rightarrow \infty)\simeq \sqrt{\lambda/2}$. 

{\bf The GJV model}

Next, a more interesting model, one that still has still conformal symmetry, but less supersymmetry, the GJV model  \cite{Guarino:2015jca}.
The gravity dual is a warped, squashed $AdS_4\times S^6$, corresponding to the fixed point of the field theory on $N$ D2-branes with 
Romans mass $m$, so the gauge group is $SU(N)$ and the gauge fields are SYM+CS, with ${\cal N}=2$ supersymmetry 
and $SU(3)\times U(1)_R$ symmetry. 

The Penrose limit was analyzed in \cite{Araujo:2017hvi}. In this case, there are 3 complex scalars, out of which one is charged under 
the symmetry corresponding to the pp wave null geodesic, $Z$, that builds the closed string vacuum, and the other two, $\phi_m$, $m=1,2$, 
are not (so $J(Z)=1/2$ and $J(\phi_m)=0$). Also we have $\Delta(Z,\bar Z,\phi_m, \bar \phi_m)=1/2$. 
The bosonic oscillators inserted into the trace $|0;p^+\rangle\sim\Tr[Z^J]$ are $\phi_m,\bar\phi_m$ for $\Delta-J=1/2$ and 
$D_\mu, \bar Z$ at $\Delta-J=1$. 

Now, since we have even less supersymmetry (though the theory is still at a conformal point), 
we have $\Delta-J=\sqrt{(\Delta-J)_{0,i}^2+f_i(\lambda)p^2}$, so for each different string oscillator (field insertion) we have
not only different BPS values $(\Delta-J)_{0,i}$, but also different 't Hooft coupling dependence $f_i(\lambda)$ in front of $p^2$. 
It is then not surprising that the same situation happens in our MNa case, just a bit more general that this. 

{\bf Phenomenological holographic cosmology}

Although a spin chain has not been described in this case, we will also consider the case of the phenomenological field theory 
model dual to holographic cosmology, defined in \cite{McFadden:2009fg}. This model was shown to match CMBR data as well as 
the standard $\Lambda$-CDM plus inflation \cite{Afshordi:2016dvb} and to also solve the usual puzzles of Big Bang cosmology
as well as inflation \cite{Nastase:2019rsn,Nastase:2020uon}, so is potentially very interesting for phenomenology. 

The three-dimensional phenomenological field theory  model is the most general $SU(N)$ gauge theory with adjoint scalars and fermions
that is superrenormalizable and has generalized conformal invariance. The latter means that the only mass scale in the theory is the 
overall coupling constant factor $\frac{1}{g^2_{YM}}$ or, equivalently, that if the theory would be derived by dimensional reduction from 
4 dimensions, the 4 dimensional theory would be conformal invariant (so that the KK reduced coupling constant factor gets a scale, 
$g^2_{YM}=g^2_{YM,4d}/(2\pi R)$. 

Because of generalized conformal invariance, the correlators of the theory will depend only on the combination $g^2_{YM}N/q$, with 
$q$ the momentum scale, and would be scale-invariant at tree level. One calculates 2-point functions of gauge-invariant 
current operators: the energy momentum tensor $T_{\mu\nu}$ and a global symmetry current $j_\mu^a$, and from their behaviour 
one extracts the anomalous dimension $\Delta$ of these operators, just like we did in the case of the BMN operators ${\cal O}_\a$
(see \cite{McFadden:2009fg,Afshordi:2016dvb,Nastase:2019rsn,Nastase:2020uon}). Besides the scale-invariant tree-level result, 
one gets loop corrections that are of the type $\frac{g^2_{YM}N}{q}\log q $, the behaviour with $\log q$ appearing since the 
loop integrals are divergent. 

We see then that this generalized conformal case 
is an intermediate situation between the conformal ABJM and GJV cases and the non-conformal MNa theory. We still have $\log q$ 
behaviour, but the coupling $g^2_{YM}$ doesn't run. In our MNa case, however, there are no divergences, but $g^2_{YM}$ runs, 
which allows us to calculate the anomalous dimension.

\section{Conclusions and open questions}

In this paper, we have revisited the Penrose limit of the MNa solution and its field theory dual. 
The pp wave obtained from the IR of the MNa solution has simple eigevalues for the string excitations. 
We have then constructed a spin chain in the dual field theory, the 6-dimensional theory on D5-branes KK expanded on 
the $S^3$ with a twist, spin chain describing "annulons" (3-dimensional equivalent of hadrons with a ring-like structure). 

We have seen that the spin chain describes well the BPS states (the states with $n_i=0$ for the string) but, while the one-loop correction 
is also universal for all the string excitations, it cannot reproduce the mixing of the states, and it gives a different behaviour 
with the coupling, and with the scale $|x|$ than in the gravity dual, hinting at a nontrivial function depending on both $g_{YM}$ and $|x|$, 
that multiplies the $p^2\sim n^2/J^2$ factor. 

We have seen that the small amount of supersymmetry and lack of conformal invariance is certainly a factor in the mismatch, 
other possible factors being: the fact that the gravity dual is expanded in the IR, but in the perturbative calculations we need to 
consider the UV at least insofar as taking $|x|\rightarrow 0$, though $g^2_{YM}N|x|$ can still be large; the fact that the null geodesic 
for the Penrose limit has a component in a non-isometric direction; and the effect of additional Feynman diagrams. 

Among possible open questions are: 

\begin{itemize}

\item The further analysis of additional Feynman diagrams, possible mixing between different insertions, and better analysis of 
the $D_{x_i}$ and $D_t$ insertions. 

\item The possibility of Penrose limits in the UV of the MNa gravity dual, and comparison with the field theory results obtained here. 

\item Comparison of this case to the Maldacena-N\'{u}\~nez case from \cite{Gimon:2002nr}.

\item The analysis of open string spin chains.

\end{itemize}


\section*{Acknowledgements}


The work of HN is supported in part by CNPq grant 301491/2019-4 and FAPESP grants 2019/21281-4 
and 2019/13231-7. HN would also like to thank the ICTP-SAIFR for their support through FAPESP grant 2016/01343-7.

\bibliography{PenroseMN}
\bibliographystyle{utphys}

\end{document}